\documentclass[11pt,a4paper]{article}
\usepackage{amsmath}
\usepackage[mathcal]{eucal}
\usepackage{latexsym}
\usepackage{amssymb}
\begin{titlepage}
\title{Instanton representation of Plebanski gravity: IX. Hamiltonian minisuperspace dynamics in undensitized momentum space variables}
\author{Eyo Eyo Ita III}
\medskip
\input amssym.def
\input amssym.tex

\def \in{\indent}

\begin{document}
\maketitle
\bigskip
\centerline{Department of Applied Mathematics and Theoretical Physics} 
\smallskip
\centerline{Centre for Mathematical Sciences, University of Cambridge, Wilberforce Road}
\smallskip
\centerline{Cambridge CB3 0WA, United Kingdom}
\smallskip
\centerline{eei20@cam.ac.uk} 

\bigskip

\begin{abstract}
In this paper we illustrate the dynamics of the instanton representation in the description of vacuum GR in minisuperspace for undensitized variables.  We uncover a new class of general solutions in both the degenerate and the nondegenerate sectors of the theory.  Additionally, the individual sectors are preserved under Hamiltonian evolution.  Finally, we present an algorithm for constructing general solutions by expansion about the isotropic sector of the instanton representation.
\end{abstract}
\end{titlepage}

\section{Introduction}

\noindent
We will acquire some intuition regarding the classical dynamics of the instanton representation of Plebanski gravity, starting in this paper with a cursory analysis of minisuperspace dynamics in undensitized momentum space variables.  The action for the instanton representation is equivalent to that in the Ashtekar variables when restricted to nondegenerate metrics.  In this paper we will examine the dynamics of both the degenerate and the nondegenerate sectors of the theory.  The instanton representation on the full phase space $\Omega_{Inst}$ is given by the first order phase space action

\begin{eqnarray}
\label{ACTIONN3}
I_{Inst}=\int{dt}\int_{\Sigma}d^3x\Bigl(\Psi_{ae}B^i_e\dot{A}^a_i+\Psi_{ae}B^i_eD_iA^a_0-N^{\mu}H_{\mu}\Bigr).
\end{eqnarray}

\noindent
$N^{\mu}=(N,N^i)$ are respectively the lapse function and the shift vector of general relativity, and $H_{\mu}=(H,H_i)$ are the Hamiltonian and the diffeomorphism constraints, given by

\begin{eqnarray}
\label{ACTIONN1}
H=(\hbox{det}B)^{1/2}\sqrt{\hbox{det}\Psi}\bigl(\Lambda+\hbox{tr}\Psi^{-1}\bigr);~~H_i=\epsilon_{ijk}B^j_aB^k_e\Psi_{ae}.
\end{eqnarray}

\noindent
The object $\Psi_{ae}$ is a $SO(3,C)\times{SO}(3,C)$ valued matrix known as the CDJ matrix \cite{NOMETRIC}, and is of mass dimension $[\Psi_{ae}]=-2$.  $B^i_e$ is the magnetic field for a self dual $SO(3,C)$ gauge connection $A^a_i$,\footnote{Our notation is that symbols from the beginning of the Latin alphabet $a,b,c,\dots$ denote internal $SO(3,C)$ indices, while from the middle of the alphabet $i,j,k,\dots$ denote spatial indices.} where

\begin{eqnarray}
\label{ACTIONN2}
B^i_a=\epsilon^{ijk}\partial_jA^a_k+{1 \over 2}\epsilon^{ijk}f_{abc}A^b_jA^c_k.
\end{eqnarray}

\noindent
Under the CDJ Ansatz

\begin{eqnarray}
\label{ACTIONN4}
\Psi^{-1}_{ae}=(\widetilde{\sigma}^{-1})^a_iB^i_e,
\end{eqnarray}

\noindent
then (\ref{ACTIONN3}) reduces for $(\hbox{det}B)\neq{0}$ and $\hbox{det}\Psi\neq{0}$ to the action for GR in the Ashtekar variables where $\widetilde{\sigma}^i_a$ is the densitized 
triad (see e.g. \cite{ASH1},\cite{ASH2},\cite{ASH3}).  Hence (\ref{ACTIONN3}) is a formulation of general relativity in which the CDJ matrix $\Psi_{ae}$ is regarded as a fundamental dynamical 
variable.\par
\indent
To study the minisuperspace dynamics of the instanton representation, we must now reduce (\ref{ACTIONN3}) to minisuperspace.  Minisuperspace is defined as the sector of the full theory where all variables are spatially homogeneous.  This means that all spatial gradients must be set to zero, which is unlike the usual definition of minisuperspace which uses Bianchi symmetry groups \cite{KODAMA}.  Hence in minisuperspace as we 
have defined it, (\ref{ACTIONN2}) reduces to

\begin{eqnarray}
\label{ACTIONN5}
B^i_a=(\hbox{det}A)(A^{-1})^i_a;~~\hbox{det}B=(\hbox{det}A)^2.
\end{eqnarray}

\noindent
We have used in (\ref{ACTIONN5}) the fact that the structure constants $f_{abc}$ for $SO(3,C)$ are numerically the same as the three dimensional epsilon symbol $\epsilon_{abc}$, in writing the determinant.  We must now reduce the constraints (\ref{ACTIONN1}) to minisuperspace.  The Hamiltonian constraint $H$ is given by

\begin{eqnarray}
\label{ACTIONN6}
H=(\hbox{det}A)\sqrt{\hbox{det}\Psi}\bigl(\Lambda+\hbox{tr}\Psi^{-1}\bigr)
\end{eqnarray}

\noindent
and the diffeomorphism constraint $H_i$ is given by

\begin{eqnarray}
\label{ACTIONN7}
H_i=(\hbox{det}A)^2(\hbox{det}A)^{-1}A^d_i\psi_d=(\hbox{det}A)A^d_i\psi_d,
\end{eqnarray}

\noindent
where $\psi_d=\epsilon_{dbf}\Psi_{bf}$ is derived from the antisymmetric part of $\Psi_{bf}$.  A direct way to obtain the Gauss' Law constraint for minisuperspace is to obtain

\begin{eqnarray}
\label{ACTIONN8}
\Psi_{ae}B^i_eD_iA^a_0=\Psi_{ae}B^i_e\bigl(\partial_iA^a_0+f^{abc}A^b_iA^c_0\bigr)=\Psi_{ae}(\hbox{det}A)(A^{-1})^i_ef^{abc}A^b_iA^c_0,
\end{eqnarray}

\noindent
and then vary (\ref{ACTIONN8}) with respect to $A^a_0$, yielding

\begin{eqnarray}
\label{ACTIONNN8}
G_a=(\hbox{det}A)\psi_a.
\end{eqnarray}

\noindent
In minisuperspace the both the Gauss' law and the diffeomorphism constraints depend linearly on $\psi_d$, and are therefore redundant.\par
\indent
The last remaining object needed is the canonical structure which determines the canonical one form, which is given by

\begin{eqnarray}
\label{ACTIONN9}
\dot{X}^{ae}=B^i_e\dot{A}^a_i=(\hbox{det}A)(A^{-1})^i_e\dot{A}^a_i.
\end{eqnarray}

\noindent
While the velocity $\dot{X}^{ae}$ is defined by (\ref{ACTIONN9}), the issue of the existence of $X^{ae}$ as a global coordinate on configuration space $\Gamma_{Inst}$ arises.\footnote{The Soo one forms $\delta{X}^{ae}$ were first introduced in  \cite{SOO} and \cite{SOO1}.}  As we will show, the existence or nonexistence of $X^{ae}$ is not relevant as far as the dynamics are concerned, since we will be able to formulate the Hamilton's equations of motion using only the velocities $\dot{X}^{ae}$ without actually making use of $X^{ae}$.\par
\indent
The organization of this paper is as follows.  Section 2 treats the nondegenerate case, computing the general solution to the equations of motion, and section 3 derives the induced time development of the spacetime metric.  Sections 4 and 5 recompute the equations of motion for the degenerate case, and construct some new solutions.  Section 6 introduces a method for constructing a general solution by expansion about the isotropic sector.

\newpage

\section{Setting the stage with nondegenerate vacuum general relativity $(X=\hbox{det}{A}\neq{0})$}

\noindent
The action for general relativity in the instanton representation in minisuperspace with Lorentzian signature is given by 

\begin{eqnarray}
\label{PURELY}
I_{Inst}=\int^T_0dt\Bigl(-i\Psi_{ae}\dot{X}^{ae}\nonumber\\
+(\hbox{det}A)\Bigl[iN\sqrt{\hbox{det}\Psi}\bigl(\Lambda+\hbox{tr}\Psi^{-1}\bigr)
+(N^iA^d_i-\theta^d)f_{dbf}\Psi_{bf}\Bigr]\Bigr),
\end{eqnarray}

\noindent 
The Hamilton's equations of motion for $\dot{X}^{ae}$ are given by

\begin{eqnarray}
\label{PURE161}
\dot{X}^{ae}=B^i_e\dot{A}^a_i={{\delta\boldsymbol{H}} \over {\delta\Psi_{ae}}}
=i\Bigl[{1 \over 2}(\Psi^{-1})^{ae}H[N]\nonumber\\
-N(\hbox{det}A)\sqrt{\hbox{det}\Psi}(\Psi^{-1}\Psi^{-1})^{ae}\Bigr]
+(\hbox{det}A)(N^iA^d_i-\theta^d)f_{dae};
\end{eqnarray}

\noindent
and for the CDJ matrix $\Psi_{ae}$ by

\begin{eqnarray}
\label{PURE162}
\dot{\Psi}_{ae}=-{{\delta\boldsymbol{H}} \over {\delta{X}^{ae}}}
=-\Bigl[i\delta_{ae}(\hbox{det}A)^{-1}H[N]\nonumber\\
+\delta_{ae}(N^iA^d_i-\theta^d)f_{dbf}\Psi_{bf}\Bigr].
\end{eqnarray}

\noindent
Since the Cauchy development of GR should be consistent with the initial value constraints, then the constraints can be applied wherever they appear in the equations of 
motion.  For $\hbox{det}B=(\hbox{det}A)^2\neq{0}$ the only constraint 
contained in (\ref{PURE161}) is the Hamiltonian constraint $H\sim{0}$.\footnote{We have used the identity $B^i_a=(\hbox{det}A)(A^{-1})^i_a$, as well as $X=\hbox{tr}X^{ae}=\hbox{det}A$, for anisotropic minisuperspace.}  On the other hand, all terms of (\ref{PURE162}) are directly proportional to constraints.\par
\indent
The diffeomorphism constraint $H_i=0$ implies that $f_{dbf}\Psi_{bf}=0$, or $\Psi_{bf}=\Psi_{(bf)}$ is symmetric in $bf$.  Application of this and the Hamiltonian constraint $H=0$ in (\ref{PURE162}) yields

\begin{eqnarray}
\label{MEET}
\dot{\Psi}_{ae}=0\rightarrow\Psi_{ae}(t)=\lambda_{(ae)}=const.~\forall{a,e}.
\end{eqnarray}

\noindent
However, for (\ref{PURE161}), only the Hamiltonian constraint can be used.  The solution to the Hamiltonian constraint is given by

\begin{eqnarray}
\label{SOLUTIONHAM}
\lambda_3=-{{\lambda_1\lambda_2} \over {\Lambda\lambda_1\lambda_2+\lambda_1+\lambda_2}}.
\end{eqnarray}

\par
\indent   
To obtain some physical insight into the configuration space dynamics, let us take the symmetric part of (\ref{PURE161}), which has the same effect as making a gauge choice $N^i=\theta^d=0$.  Then (\ref{PURE161}) 
and (\ref{PURE162}) reduce to 

\begin{eqnarray}
\label{MET3}
\dot{X}^{(ae)}=-iN(\hbox{det}A)\eta^{ae};~~\dot{\Psi}_{ae}=0,
\end{eqnarray}

\noindent
where we have defined an internal $SU(2)_{-}\otimes{S}U(2)_{-}$ metric $\eta^{ae}$ (not to be confused with the Minkoski metric $\eta^{ij}$ which has spatial indices $i,j$), by 

\begin{eqnarray}
\label{MET2}
\eta^{ae}=\sqrt{\hbox{det}\Psi}(\Psi^{-1}\Psi^{-1})^{ae}.
\end{eqnarray}

\noindent
Equation (\ref{MET3}) resembles the equations of motion for a free particle in classical mechanics, travelling ostensibly through a nine dimensional configuration space.  However, the space is actually five dimensional when restricted to the constraint surface defined by the initial value constraints.\par
\indent
Taking the trace of (\ref{MET3}) and dividing through by $X=\hbox{tr}X^{ae}=\hbox{det}A$ since $X\neq{0}$ due to nondegeneracy, we obtain\footnote{Note that $X$ is actually the Chern--Simons functional for minisuperspace, identified in \cite{SOO1} as the candidate for a time variable on configuration space in quantum cosmology.}

\begin{eqnarray}
\label{PURE164}
{{\dot{X}} \over X}={{d\hbox{ln}X} \over {dt}}=-i\eta{N}
\end{eqnarray}

\noindent
where we have defined $\eta=\delta_{ae}\eta^{ae}$ as the trace.  Since $X$ is globally a holonomic coordinate on $\Gamma_{Inst}$, then equation (\ref{PURE164}) directly integrates to

\begin{eqnarray}
\label{PURE165}
X(t)=(\hbox{det}A(t))=X_0e^{-i{\eta}\int^t_0N(t^{\prime})dt^{\prime}}
\end{eqnarray}

\noindent
where $X_0=X(0)$.  Note the mass dimension $[\eta]=1$ which cancels the negative mass dimension of time $[t]=-1$ so that the argument of the exponential is dimensionless.  Substituting (\ref{PURE165}) 
back into (\ref{MET3}) we obtain the equation for the velocity $\dot{X}^{ae}$ 

\begin{eqnarray}
\label{PURE166}
\dot{X}^{(ae)}=-iN\eta^{ae}\Bigl(X_0e^{-i{\eta}\int^t_0N(t^{\prime})dt^{\prime}}\Bigr)
={{\eta^{ae}} \over \eta}{d \over {dt}}\Bigl(X_0e^{-i{\eta}\int^t_0N(t^{\prime})dt^{\prime}}\Bigr).\nonumber\\
\end{eqnarray}

\noindent
The right hand side of (\ref{PURE166}) is a total time derivative since according to (\ref{MEET}) and (\ref{MET2}), $\eta^{ae}$ is constant in time.  This implies that for the equation to make sense the left hand side must as well be a total derivative.  However, there is no coordinate $X^{ae}$ on configuration space, since $\delta{X}^{ae}=B^i_e\delta{A}^a_i\not\subset\wedge^a(\Gamma_{Inst})$ is not an exact one form except for its trace.  However, there exist configurations for which a densitized $X^{ae}$ can in some sense be defined.  For instance, for diagonal connections $A^a_i=\delta^a_iA^a_a$ one has

\begin{eqnarray}
\label{ONEHAS}
X=(\hbox{det}A)=A^1_1A^2_2A^3_3;~~B^1_1=A^2_2A^3_3;~~B^2_2=A^3_3A^1_1;~~B^3_3=A^1_1A^2_2
\end{eqnarray}

\noindent
and (\ref{MET3}) reduces to

\begin{eqnarray}
\label{ONEHAS1}
A^2_2A^3_3\dot{A}^1_1=-iN\eta^{11}(A^1_1A^2_2A^3_3)\longrightarrow{{\dot{A}^1_1} \over {A^1_1}}=-iN\eta^{11},
\end{eqnarray}

\noindent
and likewise for $A^2_2$ and $A^3_3$.  Then (\ref{ONEHAS1}) integrates to

\begin{eqnarray}
\label{ONEHAS2}
A^f_f(t)=A^f_f(0)\Bigl({{X(t)} \over {X(0)}}\Bigr)^{\eta^{ff}/\eta},
\end{eqnarray}

\noindent
for $f=1,2,3$.  The result is that all components of the connection evolve with respect to $X$, seen as a time variable on configuration space.

\subsection{Dynamics of the spacetime metric}

Since the instanton representation is a metric-free description of gravity, then the spacetime metric $g_{\mu\nu}$ is a derived quantity.  The lapse function $N(t)$, can still be chosen arbitrarily, but the spatial 3-metric $h_{ij}$ is determined dynamically through the evolution of the instanton representation phase space.  The spatial three metric $h^{ij}$ is given by

\begin{eqnarray}
\label{MET}
h^{ij}={{\Psi_{ae}\Psi_{af}} \over {\hbox{det}\Psi}}\Bigl({{B^i_eB^j_f} \over {\hbox{det}B}}\Bigr)={{\widetilde{\sigma}^i_a\widetilde{\sigma}^j_a} \over {\hbox{det}\widetilde{\sigma}}}.
\end{eqnarray}

\noindent
The relation to $\eta_{ae}$ in (\ref{MET2}) stems more directly from the covariant form

\begin{eqnarray}
\label{MET1}
h_{ij}=(\hbox{det}\Psi)(\Psi^{-1}\Psi^{-1})^{ef}(B^{-1})^e_i(B^{-1})^f_j(\hbox{det}B),
\end{eqnarray}

\noindent
whereupon the following relation can be written

\begin{eqnarray}
\label{MOT1}
\eta^{ae}=\sqrt{\hbox{det}\Psi}(\Psi^{-1}\Psi^{-1})^{ae}\equiv(\hbox{det}\Psi)^{-1/2}h_{ij}\Bigl({{B^i_aB^j_e} \over {\hbox{det}B}}\Bigr).
\end{eqnarray}

\noindent
Making use of the minisuperspace relations $B^i_a=(\hbox{det}A)(A^{-1})^i_a$ and $\hbox{det}B=(\hbox{det}A)^2=X^2$, and $\sqrt{\hbox{det}\Psi}=(\hbox{det}\eta)^{-1}$, then (\ref{MET1}) can be written as

\begin{eqnarray}
\label{MOT}
h_{ij}\sim{h}_{ij}(t;\lambda)=(\hbox{det}\eta)^{-1}\eta^{ae}(\lambda)A^a_i(t)A^e_j(t).
\end{eqnarray}

\indent
Observe that $h_{ij}$ has acquired the label of $\eta^{ae}\in{GL}(5,C)$, and also depends on $A^a_i$.\footnote{According to the previous section, the information encoding the evolution of specific components of $A^a_i$, is tied up in the combination $\delta{X}^{ae}=B^i_e\delta{A}^a_i$.  The trace of $X^{ae}$, namely the Chern--Simons invariant, undergoes a well-defined evolution (\ref{PURE165}) from which some information regarding $A^a_i(t)$ can be inferred.}  The part of $h_{ij}$ which can always be unambigously specified is $h=\hbox{det}(h_{ij})=(\hbox{det}\eta)^{-2}X^2$, whose time evolution is given by

\begin{eqnarray}
\label{MOAT}
\sqrt{h(t)}=(\hbox{det}\eta)^{-1}X(0)e^{-i{\eta}\int^t_0N(t^{\prime})dt^{\prime}}.
\end{eqnarray}

\noindent
Hence in the general solution for the nondegenerate case even $h$, acquires the label of $\lambda_{ae}\in{GL(5,C)}$.  One could then construct a spacetime metric

\begin{eqnarray}
\label{ISOM}
ds^2=-N^2dt^2+(\hbox{det}\eta)^{-1}\eta^{ae}\boldsymbol{\omega}^a\otimes\boldsymbol{\omega}^e,
\end{eqnarray}

\noindent
whence the Ashtekar potential $A^a_i$ becomes absorbed into the definition of the one forms $\boldsymbol{\omega}^i$, given by

\begin{eqnarray}
\label{ISOM1}
\boldsymbol{\omega}^a=A^a_i(dx^i+N^idt).
\end{eqnarray}

\noindent
While (\ref{ISOM1}) resembles the invariant one forms defined for Bianchi groups \cite{KODAMA}, this is not the case since $A^a_i$ does not satisfy the Maurer--Cartan equation since it is not a flat connection.\par
\indent
For the configuration (\ref{ONEHAS2}) one obtains the time evolution for the spatial 3-metric

\begin{eqnarray}
\label{ISOM2}
h_{ij}(t)=\delta_{if}\delta_{jf}(\hbox{det}\eta)^{-1}\eta^{ff}(A^f_f(0))^2\Bigl({{X(t)} \over {X(0)}}\Bigr)^{2\eta^{ff}/\eta},
\end{eqnarray}

\noindent
which also evolves in relation to $X(t)$, seen as a time variable on configuration space.  This can be written explicitly in terms of metric variables as

\begin{eqnarray}
\label{ISOM3}
h_{ij}(t)=\delta_{if}\delta_{jf}h_{ff}(0)\Bigl({{h(t)} \over {h(0)}}\Bigr)^{\eta^{ff}/\eta},
\end{eqnarray}

\noindent
where now the determinant of the 3-metric plays the role of the time variable, and the components of the metric evolve with respect to it.  From (\ref{ISOM3}) one finds the physical interpretation of the components of 
the connection $A^f_f$ in terms of $h_{ij}$, which fixes its value at $t=0$ to

\begin{eqnarray}
\label{ISOM4}
h_{ff}(0)=(\hbox{det}\eta)^{-1}\eta^{ff}(A^f_f(0))^2.
\end{eqnarray}

\noindent
To obtain a real section of GR, one imposes reality conditions on the instanton representation variables so that the 3-metric is real.  For example, we must have $\eta^{ff}/\eta$ real for each $f$, namely that

\begin{eqnarray}
\label{ISOM5}
Im(\eta^{11}\eta^{22})=Im(\eta^{22}\eta^{33})=Im(\eta^{33}\eta^{11})=0.
\end{eqnarray}

\noindent
The real parts of $\eta^{11}$, $\eta^{22}$ and $\eta^{33}$ and one imaginary part, say $Im(\eta^{33})$ are freely specifiable, which fixes $Im(\eta^{11})$ and $Im(\eta^{22})$.  In order for the metric to be 
real at $t=0$, then (\ref{ISOM4}) in turn requires that each component of the connection be either pure real or pure imaginary.

\newpage

\section{The degenerate sector (\hbox{det}A=0)}

\noindent
Having obtained a solution for the nondegenerate sector, let us now examine the degenerate sector $(\hbox{det}A)=0$.  This implies that the metric is also degenerate, since $h=(\hbox{det}A)^2(\hbox{det}\Psi)$.  While this may be the case, as we will show, the time evolution of the CDJ matrix is still well-defined.  Recall the expression for the Hamiltonian constraint 

\begin{eqnarray}
\label{DEGEN}
H=(\hbox{det}A)\sqrt{\hbox{det}\Psi}\bigl(\Lambda+\hbox{tr}\Psi^{-1}\bigr),
\end{eqnarray}

\noindent
which includes $\hbox{det}A$ as part of its definition, therefore when $\hbox{det}A=0$ (\ref{DEGEN}) is trivially satisfied with no restrictions on $\Psi_{ae}$.  This means that the CDJ matrix $\Psi_{ae}$ is now free to evolve in time.  Let us now revisit the equations of motion (\ref{PURE161}) and (\ref{PURE162}) under the condition of degeneracy.  Starting with the equation 
for $X^{ae}$, requoting (\ref{PURE161}) for completeness,

\begin{eqnarray}
\label{DEGEN1}
\dot{X}^{ae}=B^i_e\dot{A}^a_i
=i\Bigl[{1 \over 2}(\Psi^{-1})^{ae}H[N]\nonumber\\
-N(\hbox{det}A)\sqrt{\hbox{det}\Psi}(\Psi^{-1}\Psi^{-1})^{ae}\Bigr]
+(\hbox{det}A)(N^iA^d_i-\theta^d)f_{dae}.
\end{eqnarray}

\noindent
Since $\hbox{det}A=0$, then (\ref{DEGEN1}) the trace of (\ref{DEGEN1}) vanishes.  This implies that the trace $X$ is numerically constant, and since $X(t)=\hbox{det}A=0=X_0$, it remains degenerate for all times.  Therefore, the degenerate case remains a distinct sector of vacuum GR which cannot be bridged.\footnote{Hence, the configuration space variables cannot evolve in time from the $X=0$ sector into the $X\neq{0}$ sector, since the equations of motion are assumed to hold for all time.  Consequently, topology change is precluded, unlike in \cite{DEGENERATE} and works by other authors.}\par
\indent
Moving on to the equation of motion for $\Psi_{ae}$, we have from (\ref{PURE162}) that

\begin{eqnarray}
\label{DEGEN3}
\dot{\Psi}_{ae}
=-i\delta_{ae}N\sqrt{\hbox{det}\Psi}\bigl(\Lambda+\hbox{tr}\Psi^{-1}\bigr)
-\delta_{ae}(N^iA^d_i-\theta^d)f_{dbf}\Psi_{bf}.
\end{eqnarray}

\noindent
The Gauss' law and the diffeomorphism constraints imply that $\Psi_{bf}=\Psi_{(bf)}$ is symmetric in $b,f$, therefore the second term of (\ref{DEGEN3}) vanishes and we are left with the equations

\begin{eqnarray}
\label{DEGEN66}
\dot{\Psi}_{ae}=-i\delta_{ae}N\sqrt{\hbox{det}\Psi}\bigl(\Lambda+\hbox{tr}\Psi^{-1}\bigr);~~\dot{X}^{ae}=0.
\end{eqnarray}

\noindent
Comparison of (\ref{DEGEN66}), corresponding to $(\hbox{det}A)=0$ with the analogous equations (\ref{MET3}) for $(\hbox{det}A)\neq{0}$ reveals the following contrast.  Whereas for $(\hbox{det}A)\neq{0}$ the 
momentum variable $\Psi_{ae}$ was constant in time while $\dot{X}^{ae}$ was nontrivial, we see for $(\hbox{det}A)=0$ that it is $\dot{X}^{ae}$ which vanishes while $\Psi_{ae}$ inherits a nontrivial time evolution.  The `configuration' and the `momentum' space have essentially `exchanged' roles as a consequence of the degeneracy condition.\par
\indent
Equation (\ref{DEGEN66}) states that the time derivative of $\Psi_{ae}$ is an isotropic matrix, which means that the off-diagonal parts are numerical constants.  Since $\Psi_{[ae]}=0$ on account of the kinematic constraints, we may assume that $\Psi_{ae}$ is a symmetric matrix of the form

\begin{displaymath}
\Psi_{ae}(t)=
\left(\begin{array}{ccc}
a(t) & W & V\\
W & b(t) & U\\
V & U & c(t)\\
\end{array}\right)
\end{displaymath}

\noindent
where $U$, $V$ and $W$ are arbitrary numerical constants.  Without loss of generality we can set $U=V=W=0$, since $\Psi_{ae}$ can always be diagonalized if it is nondegenerate.  Note that these initial conditions are preserved on account of the equations of motion, and the nontrivial equations of motion then reduce to

\begin{eqnarray}
\label{DEGEN9}
\dot{a}=\dot{b}=\dot{c}=iN\sqrt{abc}\Bigl(\Lambda+{1 \over a}+{1 \over b}+{1 \over c}\Bigr).
\end{eqnarray}

\noindent
The time derivates of the diagonal elements are equal, which implies that these elements must be equal, within numerical constants, to each other 

\begin{eqnarray}
\label{DEGEN10}
b(t)=a(t)+k_1;~~c(t)=a(t)+k_2
\end{eqnarray}

\noindent
for arbitrary constants $k_1$ and $k_2$.  In the general case the equation of motion can be integrated

\begin{eqnarray}
\label{DEGEN10}
\int^{a(t)}_{a_0}\bigl(a(a+k_1)(a+k_2)\bigr)^{-1/2}\Bigl(\Lambda+{1 \over a}+{1 \over {a+k_1}}+{1 \over {a+k_2}}\Bigr)^{-1}da=i\int^t_0N(t^{\prime})dt^{\prime}.
\end{eqnarray}

\noindent
The left hand side of (\ref{DEGEN10}) can be written in closed form in terms of known functions, but we do not display the result it here.  Let us rather use a compact notation to decribe the integral

\begin{eqnarray}
\label{DEGEN11}
I_{\Lambda}(a;k_1,k_2)=i\int^t_0N(t^{\prime})dt^{\prime}=\tau(t).
\end{eqnarray}

\noindent
One can then in principle invert (\ref{DEGEN11}) to find the evolution of $a=a(\tau)$ as a function of the `time' $\tau$, which depends on the choice of lapse function $N$ and is labelled by the constants $k_1$ and $k_2$.

\subsection{A few simple cases within the diagonal sector}

\noindent
Let us now illustrate a few simple examples for the degenerate case for which short expressions can be written.\par
\noindent
Case (i): Isotropic case with $\Lambda\neq{0}$.  In this case we have $k_1=k_2=0$.  The equation of motion reduces to

\begin{eqnarray}
\label{DEGEN12}
\dot{a}=iNa^{3/2}\Bigl(\Lambda+{3 \over a}\Bigr)
\end{eqnarray}

\noindent
Equation (\ref{DEGEN12}) directly integrates to

\begin{eqnarray}
\label{DEGEN13}
I_{\Lambda}(a;0,0)=\tau(t);~~
a(\tau)={3 \over {\Lambda}}\hbox{tan}\Bigl[\hbox{tan}^{-1}\Bigl(\sqrt{{{\Lambda{a}_0} \over 3}}+{{\sqrt{3\Lambda}} \over 2}\tau\Bigr)\Bigr]
\end{eqnarray}

\noindent
For $\Lambda=0$ we have

\begin{eqnarray}
\label{DEGEN14}
a(\tau)=\Bigl(\sqrt{a_0}+{3 \over 2}\tau\Bigr)^2.
\end{eqnarray}

\par
\noindent
Case (ii): One diagonal degree of freedom.  In this case $k_2=0$, $k_1=k$ where $k$ is an arbitrary numerical constant.  The we have

\begin{eqnarray}
\label{DEGEN15}
\dot{a}=iNa\sqrt{a+k}\Bigl(\Lambda+{2 \over a}+{1 \over {a+k}}\Bigr)
\end{eqnarray}

\noindent
Equation (\ref{DEGEN15}) integrates to

\begin{eqnarray}
\label{DEGEN17}
I_{\Lambda}(a;k,0)
=\sqrt{{2 \over {\Lambda}}}\Biggl[{{r_{-}\hbox{tanh}^{-1}\Bigl({{\sqrt{2\Lambda(a+k)}} \over {r_{-}}}\Bigr)
-r_{+}\hbox{tanh}^{-1}\Bigl({{\sqrt{2\Lambda(a+k)}} \over {r_{+}}}\Bigr)\biggr)} \over {k\Lambda-3-r_{-}^2}}\Biggr]
=\tau
\end{eqnarray}

\noindent
where we have defined the dimenionless constants

\begin{eqnarray}
\label{DEGEN18}
r_{-}=\sqrt{k\Lambda-3-\sqrt{k^2{\Lambda}^2-2k\Lambda+9}};~~
r_{+}=\sqrt{k\Lambda-3+\sqrt{k^2{\Lambda}^2-2k\Lambda+9}}.
\end{eqnarray}

\noindent
While solved in closed form, (\ref{DEGEN18}) is nontrivial to invert for $a=a(t)$, but nevertheless the solution is implicit.  The $\Lambda=0$ case leads to the relation

\begin{eqnarray}
\label{DEGEN18}
I_0(a;k,0)={2 \over 3}\Bigl[\sqrt{a+k}-\sqrt{{k \over 3}}\hbox{tanh}^{-1}\Bigl(\sqrt{3\Bigl(1+{a \over k}\Bigr)}\Bigr)=\tau.
\end{eqnarray}

\noindent
Likewise, the relation (\ref{DEGEN18}) is implicit but still nevertheless illustrates the integrability of the system.\par
\noindent
Case (iii): General diagonal case for $\Lambda=0$.  This is given by

\begin{eqnarray}
\label{DEGEN19}
I_0(a;k_1,k_2)
=\int_{a_0}^{a(t)}{{\sqrt{a(a+k_1)(a+k_2)}} \over {3a^2+2a(k_1+k_2)+k_1k_2}}da=\tau
\end{eqnarray}

\noindent
The relation (\ref{DEGEN19}) can as well be integrated in closed form in terms of known functions, though we do not display the final expression here.

\newpage

\section{The degenerate general case}

\noindent
We now treat the general solution for the degenerate case, solving the equation of motion

\begin{eqnarray}
\label{DEGEN6}
\dot{\Psi}_{ae}=-i\delta_{ae}N\sqrt{\hbox{det}\Psi}\bigl(\Lambda+\hbox{tr}\Psi^{-1}\bigr);~~\dot{X}^{ae}=0
\end{eqnarray}

\noindent
for a more general form of $\Psi_{ae}$.  This is given by

\begin{displaymath}
\Psi_{ae}=
\left(\begin{array}{ccc}
a & W & V\\
W & b & U\\
V & U & c\\
\end{array}\right).
\end{displaymath}

\noindent
It will be convenient to parametrize $\Psi_{ae}$ by its isotropic and its non-isotropic parts, as in

\begin{eqnarray}
\label{KLU}
\Psi_{ae}=\delta_{ae}\varphi+\epsilon_{ae}.
\end{eqnarray}

\noindent
In (\ref{KLU}), $\varphi$ is the isotropic part and $\epsilon_{ae}$ is defined as the `CDJ deviation matrix'.  Substitution of (\ref{KLU}) into (\ref{DEGEN6}) yields

\begin{eqnarray}
\label{DEGEN77}
\dot{\epsilon}_{ae}=0;~~\epsilon_{ae}=const.~\forall{a,e}
\end{eqnarray}

\noindent
The deviation matrix $\epsilon_{ae}$ is in general an 8 by 8 matrix of arbitrary complex numerical constants.\par
\indent
Let us now re-examine the equation of motion for $\Psi_{ae}$.  This can be written as one equation labelled by the constants $\epsilon_{ae}\in{GL}(8,C)$.

\begin{eqnarray}
\label{KALOO}
\dot{\varphi}=-iN\sqrt{\hbox{det}(\delta_{ae}\varphi+\epsilon_{ae}})\Bigl(\Lambda+\hbox{tr}(\delta_{ae}\varphi+\epsilon_{ae})^{-1}\Bigr).
\end{eqnarray}

\noindent
To solve the equation of motion (\ref{KALOO}) we will need the determinant, given by

\begin{eqnarray}
\label{KLU1}
\hbox{det}\Psi_{ae}=\hbox{det}(\delta_{ae}\varphi+\epsilon_{ae})
=\varphi^3+\varphi^2\epsilon+{1 \over 2}\varphi{Var}\epsilon+\hbox{det}\epsilon
\end{eqnarray}

\noindent
where we have defined

\begin{eqnarray}
\label{KLU1}
\epsilon=\delta_{ae}\epsilon_{ae}=\hbox{tr}(\epsilon_{ae});~~
Var\epsilon=(\delta_{bf}\delta_{cg}-\delta_{cf}\delta_{bg})\epsilon_{bf}\epsilon_{cg}=(\hbox{tr}\epsilon)^2-\hbox{tr}\epsilon^2.
\end{eqnarray}

\noindent
We will also need the inverse of (\ref{KLU}).  Let us assume for simplicity that the isotropic part $\varphi$ is the largest component.\footnote{The trace is the dynamical component and will vary in time.  Therefore, the expansion is good only for $t=\tau$ such that $\varphi(\tau)>\hbox{max}\{\epsilon_{ae}\}$.  Since the non-isotropic components are numerical constants, then $\forall~t$ such that $\varphi(t)>\epsilon_{ae}~\forall{a,e}$ do not hold, then a singularity exists and one can must attempt to find the solution for different $\epsilon_{ae}$.  Hence we have shown that singularities can develop only when considering degenerate 
configurations.}  Hence, $\varphi>\epsilon_{ae}~\forall{a,e}$, with $\varphi\neq{0}$.  Then we have

\begin{eqnarray}
\label{KLU2}
\Psi^{-1}_{ae}=\bigl(\delta_{ae}\varphi+\epsilon_{ae}\bigr)^{-1}
=\varphi^{-1}\delta_{aa_0}\Bigl(\sum_{n=0}^{\infty}(-1)^n\varphi^{-n}\epsilon_{a_0a_1}\epsilon_{a_1a_2}\dots\epsilon_{a_{n-1}a_n}\Bigr)\delta_{a_ne}
\end{eqnarray}

\noindent
The trace of (\ref{KLU2}) is given by

\begin{eqnarray}
\label{KLU3}
\hbox{tr}\Psi^{-1}
=\varphi^{-1}\sum_{n=0}^{\infty}(-1)^n\varphi^{-n}\epsilon_{aa_1}\epsilon_{a_1a_2}\dots\epsilon_{a_{n-1}a}
\end{eqnarray}

\noindent
Likewise, the square root of the determinant can be expanded as in

\begin{eqnarray}
\label{KLU4}
\sqrt{\hbox{det}\Psi}=\varphi^{3/2}\Bigl(1+{\epsilon \over \varphi}+{{Var\epsilon} \over {2\varphi^2}}+{{\hbox{det}\epsilon} \over {\varphi^3}}\Bigr)^{1/2}
\end{eqnarray}

\noindent
Then one can use the infinite binomial series expansion

\begin{eqnarray}
\label{KLU5}
\Bigl(1+{\epsilon \over \varphi}+{{Var\epsilon} \over {2\varphi^2}}+{{\hbox{det}\epsilon} \over {\varphi^3}}\Bigr)^{1/2}
=\sum_{n=0}^{\infty}{{(-1)^n(2n)!} \over {(1-2n)n!^24^n}}\Bigl({\epsilon \over \varphi}+{{Var\epsilon} \over {2\varphi^2}}+{{\hbox{det}\epsilon} \over {\varphi^3}}\Bigr)^n.
\end{eqnarray}

\noindent
We must now put (\ref{KLU3}) and (\ref{KLU5}) back into (\ref{KALOO}).

\subsection{Asymptotic expansion about the isotropic sector}

\noindent
We are now ready to write down the general solution for the degenerate case to all orders.  The equation of motion for $\varphi$ can be written as

\begin{eqnarray}
\label{ASSYM}
\Bigl(\Lambda+{3 \over \varphi}\Bigr)\varphi^{-3/2}\dot{\varphi}
=-iN^{\prime}E(\varphi,\epsilon)
\end{eqnarray}

\noindent
where we have defined a `correction' factor for non-isotropy $E$

\begin{eqnarray}
\label{ASSYM1}
E(\varphi,\epsilon)=\biggl[1-\varphi^{-2}\Bigl(\Lambda+{3 \over \varphi}\Bigr)^{-1}I_1(\varphi,\epsilon_{ae})\biggr]I_3(\varphi,\epsilon_{ae}),
\end{eqnarray}

\noindent
where 

\begin{eqnarray}
\label{ASSYM2}
I_1(\varphi,\epsilon_{ae})=\delta_{a_0}\sum_{n=1}^{\infty}(-1)^n\varphi^{-n}\epsilon_{aa_1}\epsilon_{a_1a_2}\dots\epsilon_{a_{n-1}a_n}\delta_{a_n,e};\nonumber\\
I_3(\varphi,\epsilon_{ae})=
\Bigl(1+{\epsilon \over \varphi}+{{Var\epsilon} \over {2\varphi^2}}+{{\hbox{det}\epsilon} \over {\varphi^3}}\Bigr)^{1/2}
\end{eqnarray}

\noindent
The left hand side of (\ref{ASSYM}) reduces to the isotropic case considered in (\ref{DEGEN12}).  Upon integration, we have

\begin{eqnarray}
\label{ASSYM3}
\int^t_0\Bigl(\Lambda+{3 \over \varphi}\Bigr)\varphi^{-3/2}d\varphi
=-i\int_0^tN(t^{\prime})E(\varphi(t^{\prime}),\epsilon)dt^{\prime}
\end{eqnarray}

\noindent
which can be written in the form

\begin{eqnarray}
\label{ASSYM4}
\hbox{tan}^{-1}\Bigl(\sqrt{{{\Lambda\varphi(t)} \over 3}}\Bigr)=
\hbox{tan}^{-1}\Bigl(\sqrt{{{\Lambda\varphi_0} \over 3}}\Bigr)
+{{\sqrt{3\Lambda}} \over 2}\int^t_0E(\varphi(t^{\prime}),\epsilon)dt^{\prime}.
\end{eqnarray}

\noindent
Equation (\ref{ASSYM4}) further simplifies to

\begin{eqnarray}
\label{ASSYM31}
\sqrt{\varphi(t;\epsilon)}=\Bigl(\sqrt{\varphi_0}+{3 \over {\Lambda\sqrt{\varphi_0}}}\Bigr)\Bigl[1-\sqrt{{{\Lambda\varphi_0} \over 3}}
\hbox{tan}{{\sqrt{3\Lambda}} \over 2}\int^t_0E(\varphi;\epsilon)dt^{\prime}\Bigr]^{-1}-{3 \over {\Lambda\sqrt{\varphi_0}}}.
\end{eqnarray}

\noindent
We can now generate an asymptotic expansion about the exact isotropic solution by fixed point iteration procedure, assuming that the right hand side of (\ref{ASSYM3}) is small.  First, choose  a particular state $\epsilon_{ae}$, thought of as a constant vector in an eight complex dimensional space.  Next, define a sequence of functions of time $\varphi_n(t)$, such that $\varphi=\varphi_0$, which is the initial value of the isotropic part of the CDJ matrix at time $t=0$.\footnote{This time labels the beginning of the universe, and $\varphi_0$ is an essential parameter of the theory.}  Define sequence $\varphi_n$, such that

\begin{eqnarray}
\label{ASSYM5}
\varphi_{n+1}(t)=\sqrt{{3 \over {\Lambda}}}\hbox{tan}\Bigl[\hbox{tan}^{-1}\Bigl(\sqrt{{{\Lambda\varphi_0} \over 3}}\Bigr)+
{{\sqrt{3\Lambda}} \over 2}\int^t_0E(\varphi_n(t^{\prime}),\epsilon)dt^{\prime}\Bigr].
\end{eqnarray}

\noindent
The full solution, assuming convergence of the sequence (\ref{ASSYM5}), is given by $\varphi(t)=\hbox{lim}_{n\rightarrow\infty}\varphi_n(t)$.  Hence, if it is indeed the case that there was a phase of the universe in which degenerate metrics exist, (i) The universe cannot transition into a nondegenerate state. (ii) The existence of such a configuration is labelled by eight arbitrary complex constants for the $\epsilon_{ae}$, each of which defines a state of the universe. (iii) The convergence of the sequence (\ref{ASSYM5}) is a necessary, but not sufficient condition for the well-definedness of this configuration.\footnote{There is clearly a large category of cosmological evidence that dictates that the spacetime metric today is nondegenerate.}  This in turn depends upon the initial value of $\varphi_0$ in relation to the constants $\epsilon_{ae}$.  The theorist has free 
choice of $\epsilon_{ae}$, within the range of parameters yielding a fixed point.  However, $\varphi_0$ is determined by the initial conditions of the universe, which can only be extrapolated from the cosmological data in existence today.

\newpage

\section{Conclusion}

\noindent
We have illustrated some of the dynamics of gravity in the instanton representation in minisuperspace for undensitized CDJ matrix $\Psi_{ae}$, seen as a momentum space variable.  Starting from this representation, which is defined independently of the existence of a spacetime metric, we have derived solutions which include a class of inflating spacetimes labelled by the eigenvalues of $\Psi_{ae}$.  We have also demonstrated that the big bang singularity can be avoided in this representation independently of any quantum corrections by applying the equations of motion to arbitrarily small times, including the beginning of the universe when quantum effects are presumably dominant.  We have also illustrated the dynamics of the theory in the degenerate sector, displaying an array of new solutions.  Another result of the intial value constraints development in these variables is that an initially degenerate solution will remain degenerate for all times, which eliminates topology changing configurations classically.\par
\indent
The present paper has demonstrated the anisotropic minisuperspace sector of the model.  We have also provided a prescription for obtaining general solutions by asymptotic expansion about an isotropic solution.  It is hoped that this algorithm can be of use as well in numerical treatments of general relatity.  While the configuration space variables $X^{ae}$ which would ordinarily be conjugate to $\Psi_{ae}$ are not globally integrable for generic configurations, there is certain information which can be inferred from specific combinations of these variables which corresponds to a well-defined time evolution.  A future paper along these lines will investigate the structure of $X^{ae}$, using a version of the instanton representation where the variables are densitized, and as well carry out the analogous computations of the present paper.

\end{document}